\documentclass[english,aip,manuscript,jcp,numerical,reprint]{revtex4-1}
\usepackage[T1]{fontenc}
\usepackage[utf8]{inputenc}
\usepackage{babel}
\usepackage{float}
\usepackage{booktabs}
\usepackage{amstext}
\usepackage{amssymb}
\usepackage{graphicx}
\usepackage{subscript}
\usepackage[unicode=true,pdfusetitle,
 bookmarks=true,bookmarksnumbered=false,bookmarksopen=false,
 breaklinks=false,pdfborder={0 0 0},backref=false,colorlinks=false]
 {hyperref}

\makeatletter

\date{\today}
\usepackage{natmove}

\makeatother

\begin{document}

\title{Hydrogen adsorption in metal-organic frameworks:\\
 the role of nuclear quantum effects}

\author{Mohammad~Wahiduzzaman}

\affiliation{School of Engineering and Science, Jacobs University Bremen, D-28759 Bremen,
Germany}

\author{Christian~F.~J.~Walther}

\affiliation{School of Engineering and Science, Jacobs University Bremen, D-28759 Bremen,
Germany}

\author{Thomas~Heine}

\email[To whom correspondence should be addressed. Electronic mail:~]{t.heine@jacobs-university.de}

\affiliation{School of Engineering and Science, Jacobs University Bremen, D-28759 Bremen,
Germany}

\begin{abstract}
The role of nuclear quantum effects on the adsorption of molecular hydrogen
in metal-organic frameworks (MOFs) has been investigated on grounds of
Grand-Canonical Quantized Liquid Density-Functional Theory (GC-QLDFT) calculations.
For this purpose, we have carefully validated classical H$_2$-host
interaction potentials that are obtained by fitting Born-Oppenheimer ab
initio reference data. The hydrogen adsorption has first been assessed classically
using Liquid Density-Functional Theory (LDFT) and the Grand-Canonical
Monte Carlo (GCMC) methods. The results have been compared against the
semi-classical treatment of quantum effects by applying the Feynman-Hibbs
correction to the Born-Oppenheimer-derived potentials, and by explicit treatment
within the Grand-Canonical Quantized Liquid Density-Functional Theory (GC-QLDFT).

The results are compared with experimental data and indicate pronounced
quantum and possibly many-particle effects. After validation calculations
have been carried out for IRMOF-1 (MOF-5), GC-QLDFT is applied to study
the adsorption of H$_2$ in a series of MOFs, including IRMOF-4,
-6, -8, -9, -10, -12, -14, -16, -18 and MOF-177. Finally, we discuss the
evolution of the H$_2$ quantum fluid with increasing pressure
and lowering temperature. 
\end{abstract}

\pacs{ 02.70.Uu, 05.30.-d, 05.30.Jp, 05.70.Ce, 64.30.Jk,  68.43.-h, 88.30.-k, 88.30.R- }

\keywords{QLDFT, GCMC, MOFs, H$_2$ adsorption, quantum effects}

\maketitle

\section{Introduction}

A CO$_2$ neutral economy requires efficient technologies
for the storage and transport of energy carries. Chemical storage is the
most likely solution of this problem, as we know no other carrier that
can reversibly store and release energy in similar quantity. Molecular
hydrogen (H$_2$) is the chemical system with the highest
energy per mass unit. Moreover, it can be produced by water splitting,
and its oxidation product is environmentally benign. Hydrogen already plays
an important role in the development of alternative transport systems.
It is, however, limited by the difficulty to generate a hydrogen transport
system that has a sufficient gravimetric and volumetric storage capacity.
One of the promising hydrogen storage systems that have been discussed
extensively are metal-organic frameworks (MOFs). MOFs are materials with very large internal surface area,
and thus they are capable of adsorbing significant amounts of H$_2$
at liquid nitrogen temperature ($T=77$ K) and moderate pressures~\cite{Suh2012}.

In nanoporous materials such as MOFs, H$_2$ is adsorbed by
physisorption, that is by weak, attractive interactions. Depending on the
chemical nature of the pore surfaces we distinguish between strong adsorption
sites, where a charge center at the MOF induces a dipole moment in the
adsorbed H$_2$ molecule and attracts it by up to $\sim$40
kJ~mol$^{-1}$, and weak adsorption on the surface of the
framework, with typical interaction energies in the 3-6 kJ mol$^{-1}$
range. The strong adsorption sites can be occupied at high temperature,
while the weak ones are only attractive to hydrogen molecules at liquid
nitrogen temperatures and below~\cite{Oh2014}. As the number of strong
attraction sites is limited by the framework topology and -- due to the
incorporation into the framework -- cover only a relatively small volume
of the pores, their presence is not sufficient to allow for sizeable
room temperature adsorption. For low temperatures, however, they are responsible
for a relatively high local density of the liquid hydrogen fluid, but the
main amount of adsorbed hydrogen is found to be rather homogeneously distributed
over the pore surface. Thus, at low temperature the internal surface area
of a MOF is the determining factor of its hydrogen storage capacity~\cite{Hirscher2010a}.

A large number of MOF structures have been synthesized and studied in the
last decade~\cite{Eddaoudi2002,Chae2004,Furukawa2010,Farha2010}, some
of them show remarkable surface areas and hydrogen storage capacities~\cite{Han2009,Getman2012,Suh2012}.
One of the most notable examples is MOF-210 with 6240 m$^2$
g$^{-1}$ BET surface area, which can adsorb 176 mg g$^{-1}$
H$_2$ at 77 K~\cite{Furukawa2010}. In this work we explore
the nature of the adsorbed hydrogen fluid by calculations that consider
explicitly its quantum mechanical nature. In order to achieve this goal,
we have carefully validated our computational approach. First, available
force fields (FF) to describe the interaction between H$_2$
and the MOF framework have been compared. Based on extensive validation
calculations, we have selected the FF of Fu et al.~\cite{Fu2009},
a force-field that has been parametrized from ab initio calculations and
that resembles the Born-Oppenheimer (BO), that is classical interaction
between hydrogen and the framework. Quantum effects can then be included
either by semi-classical correction (Feynman-Hibbs (FH))~\cite{Feynman1965},
or directly included into the Hamiltonian as kinetic energy operator~\cite{Patchkovskii2009,Martinez-Mesa2012,Walther2013}.
We further benchmark our approach and assess the different approximations
by their influence on the adsorption properties of the H$_2$
quantum fluid. Finally, after careful validation on MOF-5, we apply our
method, the Grand-Canonical Quantized Liquid Density-Functional Theory
(GC-QLDFT) to a series of MOFs.

\section{Quantized Liquid Density-Functional Theory}

Realistic theoretical simulations of H$_2$ adsorption at
high densities are essential in guiding the design of hydrogen adsorbent
materials. In principle, a wide range of statistical mechanics based simulation
techniques is available for this purpose, including Grand Canonical Monte
Carlo (GCMC)\cite{Frenkel2002}, molecular dynamics (MD), and classical
Density-Functional Theory of Liquids (LDFT)~\cite{Saam1977}. In most cases,
quantization of nuclear motion of the hydrogen guest molecules is 
not explicitly considered in the simulations. Although quantum effects can
be treated rigorously with path-integral (PI) techniques, e.g. in PI-GCMC
\cite{Wang1997}, such simulations are quite elaborate and are not always
trivial to interpret. Instead, it is common to semi-classically include
quantum effects in the interaction potential, either by using the Feynman-Hibbs
correction~\cite{Feynman1965}, or by using force fields that have been
optimized to match experimental results~\cite{Yang2005}.

In 2009, Patchkovskii and Heine suggested to include quantum effects into
LDFT. In the Quantized LDFT (QLDFT), they employed the Kohn-Sham technique
to introduce an effective non-interacting system of one-particle wave functions,
for which the kinetic energy operator is explicitly known~\cite{Patchkovskii2009}.
For liquid nitrogen temperatures (77 K) and above, the particles can be
described equally well using a Stefan-Boltzmann or Bose-Einstein statistics
\cite{Martinez-Mesa2011}. Many-particle effects are implicitly included
in the excess, or exchange-correlation functional (XC)~\cite{Patchkovskii2009,Martinez-Mesa2012}. As common in electronic
DFT, the XC functional also accounts for any further approximations and
models, and two XC approximations have been suggested: the local density
approximation (LDA), and the non-local weighted density approximation (WDA).
By construction, QLDFT is exact for the uniform hydrogen gas. The implementation
of QLDFT takes advantage of the sparsity of the Hamiltonian at higher 
temperature~\cite{Patchkovskii2007}. For lower temperatures, the sparsity is strongly
reduced, resulting in an increase of memory and computational requirements.
In these cases, QLDFT is not computationally competitive, and consequently
only few applications have been reported so far. The QLDFT method has been
successfully applied in the hydrogen adsorption studies in carbon foams~\cite{Martinez-Mesa2012},
porous aromatic frameworks (PAFs)~\cite{Lukose2012}
and to determine the selectivity of D$_{2}$ vs. H$_2$ adsorption
in the metal-organic framework MFU-4~\cite{Teufel2012}.

A low-temperature alternative to the Kohn-Sham scheme of QLDFT has been
suggested recently~\cite{Walther2013}: due to the bosonic nature of molecular
hydrogen, at low temperatures the lowest eigenstate dominates the occupation
of the Kohn-Sham Hamiltonian. Thus, the quantum-mechanical problem was
reformulated within the Hohenberg-Kohn-Mermin formalism of a Grand Canonical
(GC) ensemble. 

In GC-QLDFT, we directly minimize the density. However, we introduce a pseudo-wavefunction
that satisfies the equation $\rho(\vec{r})=\tilde{\psi}^2(\vec{r})$.
The main purpose of this wave function is to be able to approximate the kinetic energy operator,
which we can now conveniently write within the effective one particle approximation~\cite{Walther2013}:
\begin{equation}
T[\rho]=-\frac{\hbar^2}{2 m} \int d^3 r \tilde{\psi}\vec{\nabla}^2\tilde{\psi}.
\end{equation}
The GC-QLDFT free energy functional is then given by:
\begin{equation}
F[\rho]=-\frac{\hbar^2}{2m} \int d^3 r \tilde{\psi}\vec{\nabla}^2\tilde{\psi}+F_{\mathrm{H}}[\rho]+F_{\mathrm{xc}}[\rho]+\int d^3 r \,v_{\mathrm{ext}}(\vec{r}) \rho (\vec{r}), 
\label{[GCQLDFTFreeEnergy]}
\end{equation}
where $F_{\mathrm{H}}$ is the Hartree term, $F_{\mathrm{xc}}$ the excess correction and $v_{\mathrm{ext}}$ denotes the external potential. 

Quantum effects are thus included explicitly in the kinetic energy term which is responsible that the density vanishes
correctly close to regions with a huge external potential. Quantum effects are also included approximatively
in $F_{\mathrm{xc}}$, which is constructed to match the experimental data.\cite{Walther2013}

It is worth noting that it is difficult to compare classical and quantized LDFT, as the QLDFT functional was constructed
to fulfil the uniform fluid limit exactly, realized by including the experimental free energy
explicitly in $F_{\mathrm{xc}}$, while the classical theory starts from the Helmholtz free energy of the ideal gas
and includes additional corrections. Both methods typically
include the weighted density $\overline{\rho}$ in the excess functional.  

As in other DFT schemes, QLDFT strongly benefits from removal of systematic errors which is achieved
by requesting that experimental reference data such as the uniform fluid limit is matched exactly.
This way, various approximations that are made for the sake of numerical simplicity are compensated to 
a large extent. This includes the shapeless particle approximation, 
the one-effective-particle approximation for the kinetic energy, and even many-body effects between the particles.

The resulting grand potential is minimized by employing the Car-Parrinello technique~\cite{Car1985}, allowing for
an efficient treatment of low-temperature systems, even if the host structure
requires a very large unit cell for its proper description~\cite{Walther2013}.
In this work, we employ this variant of QLDFT within
the GC ensemble (GC-QLDFT). In order to avoid long abbreviations, we
will label this method as QLDFT in the remainder of this work.

\begin{figure*}
\centering{}\includegraphics[width=0.8\textwidth]{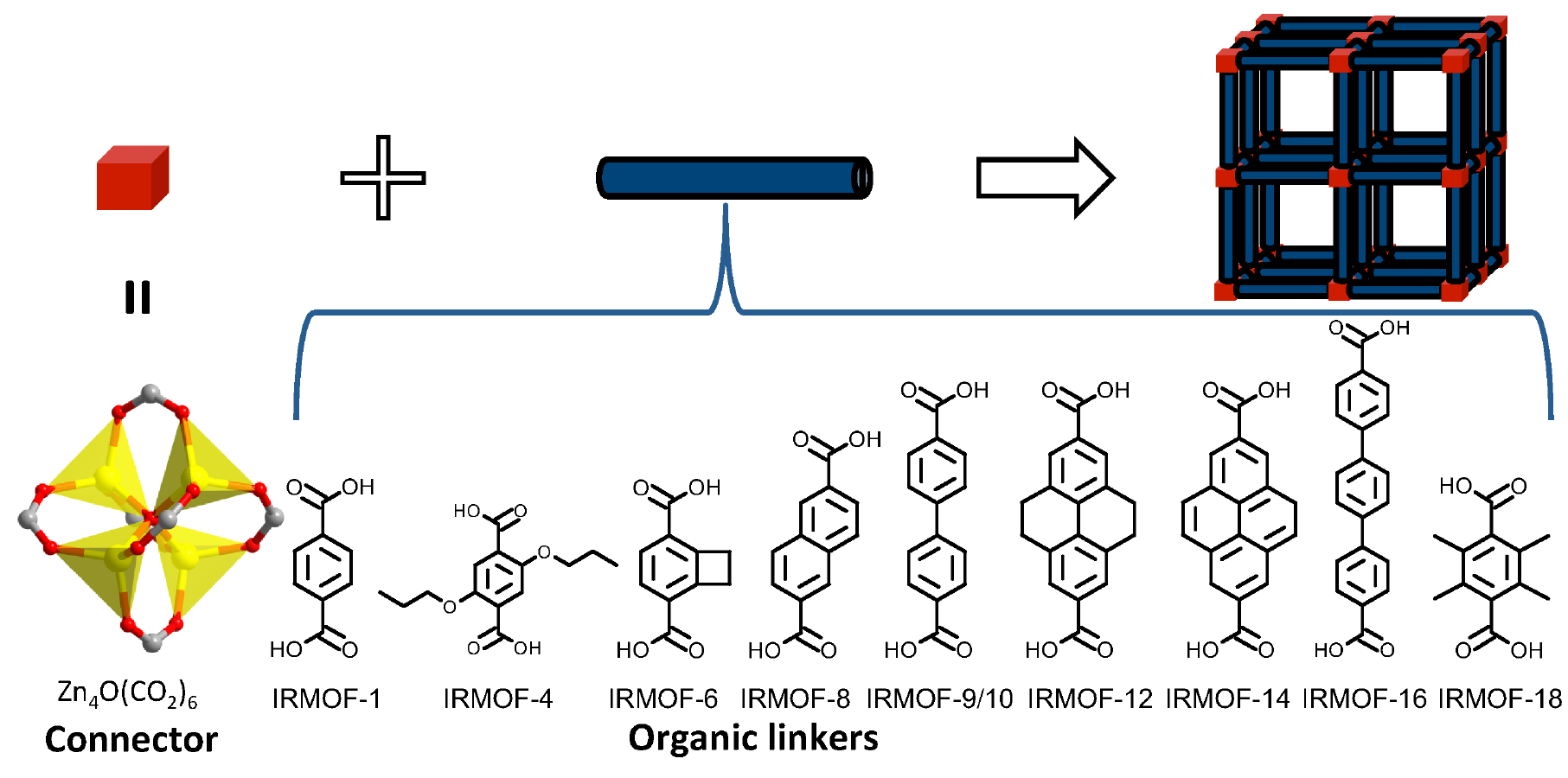}\caption{\label{fig:fig_01}Schematic representation
 of the structure topology of Zn$_4$O(CO$_2$)$_6$-based
IRMOFs. }
\end{figure*}

\section{Computational methods and models of the materials}

GCMC simulations of hydrogen adsorption were performed with the multipurpose
simulation code Music~\cite{Gupta2003}. The sorbent model was described
by a finite three-dimensional periodic $2\times2\times2$ supercell. A
tail cutoff of 12.8 Å was applied to calculate pair-wise van-der-Waals
interactions. Each Monte Carlo (MC) move consisted of an insertion attempt
of a new molecule, the deletion or translation of an existing molecule.
The probability of each of these moves were equally weighted. Since the
H$_{2}$ molecules have been modeled as shapeless particles, rotation and
intramolecular moves were neglected. For each point on the isotherm, three
million MC steps were carried out to equilibrate the system, and a further
two million MC steps were performed to calculate the ensemble average.
The non-ideal behavior of adsorption isotherms at high pressure have been
corrected by using the gas-phase fugacities calculated with Leachman et
al.'s~\cite{Leachman2009} equation of states (EOS). Quantum effects were
included by correcting the H$_2$-H$_2$ and
H$_2$-host potentials using the Feynman-Hibbs (FH)~\cite{Feynman1965}
formula:
\begin{equation}
V_{\mathrm{FH}}(r)=V(r)+\frac{\hbar^{2}}{24\mu k_{B}T}\nabla^{2}V(r),
\label{eq:FH}
\end{equation}
where $V(r)$ is the classical potential energy function, $\mu$ is the
reduced mass of interacting particles, and $k_{B}$ is the Boltzmann constant.

The QLDFT calculations have been carried out within the qLIE1 approximation,
thus including many-body interparticle interactions and quantum effects
implicitly through the excess functional~\cite{Walther2013}. QLDFT calculations
use a grid spacing of 0.2 Å and a potential cutoff of 5000 K. For the classical
LDFT variant, the cLIE XC functional has been employed, while all other
simulation parameters are identical to QLDFT.

One of the results of our simulations is the total amount of adsorbed H$_{\text{2}}$,
that is the total number of adsorbed molecules $N_{\mathrm{tot}}$ present
in the simulation box. In experiment, the excess amount adsorbed is measured.
In order to compare our results with experiment, we define the excess adsorption
as 
\begin{equation}
N_{\mathrm{ex}}=N_{\mathrm{tot}}-\rho_{\mathrm{g}}(T,P)V_{\mathrm{free}},
\label{eq:N_ex}
\end{equation}
where $\rho_{\mathrm{g}}(T,P)$ is the bulk gas density at given temperature and pressure
obtained from Leachman et al.'s~\cite{Leachman2009} EOS, 
and $V_{\mathrm{free}}$ is the H$_2$ accessible free
volume. This latter quantity is calculated as the volume within the unit
cell where the potential acting on a single H$_{2}$ molecule is less than
5000 K. 

The solvent accessible surface areas (SASA) were calculated by
rolling a probe molecule along the van-der-Waals spheres of the atoms of the MOF structure~\cite{Duren2007}.
This type of surface area is resulting from the center of the probe molecule,
i.e., the surface corresponds to an extended sphere of each atom with a
distance equal to the atom radius plus the probe radius from the center
of the atoms. A probe molecule with a diameter of 2.96 Å was used which
is equal to the Lennard-Jones $\sigma$ parameter of H$_2$
molecule and commonly used in other H$_2$ molecular simulations~\cite{Frost2006,Buch1994}.
We employed the VMD molecular visualization software~\cite{Humphrey1996,Varshney1994}
for the solvent accessible surface calculation.

Isoreticular MOFs (IRMOFs) are a series of molecular frameworks having
identical structural topology. The members of the IRMOF series share the
same connecting SBU, but different organic linkers provide a systematic
variation of pore size. The series of isoreticular MOFs studied in this
work are IRMOF-n with n=1, 4, 6, 8, 9, 10, 12, 14, 16, 18 is derived from
linking the octahedral zinc acetate unit Zn$_4$O(CO$_2$)$_6$,
with a variety of linear ditopic carboxylates~\cite{Eddaoudi2002,Rowsell2004}
(see Figure~\ref{fig:fig_01}). MOF-177 is constructed from the same connecting SBU and
the tritopic BTB linker~\cite{Chae2004}. All the IRMOFs except IRMOF-4,
are adopted from the work of Kuc et al.,~\cite{Kuc2007a} who have optimized
the structures by density-functional based tight-binding (DFTB) calculations.
The IRMOF-4 structure was constructed from the reported crystal structure
data~\cite{Eddaoudi2002}. The MOF-177 crystal structure was obtained from
Lukose et al.,~\cite{Lukose2012a} again resembling DFTB optimized structures.
The structural properties of these MOFs are given in Table~\ref{tab:table_1}.

\section{The free hydrogen gas}

\begin{figure}[h]
\centering{}\includegraphics[width=0.45\textwidth]{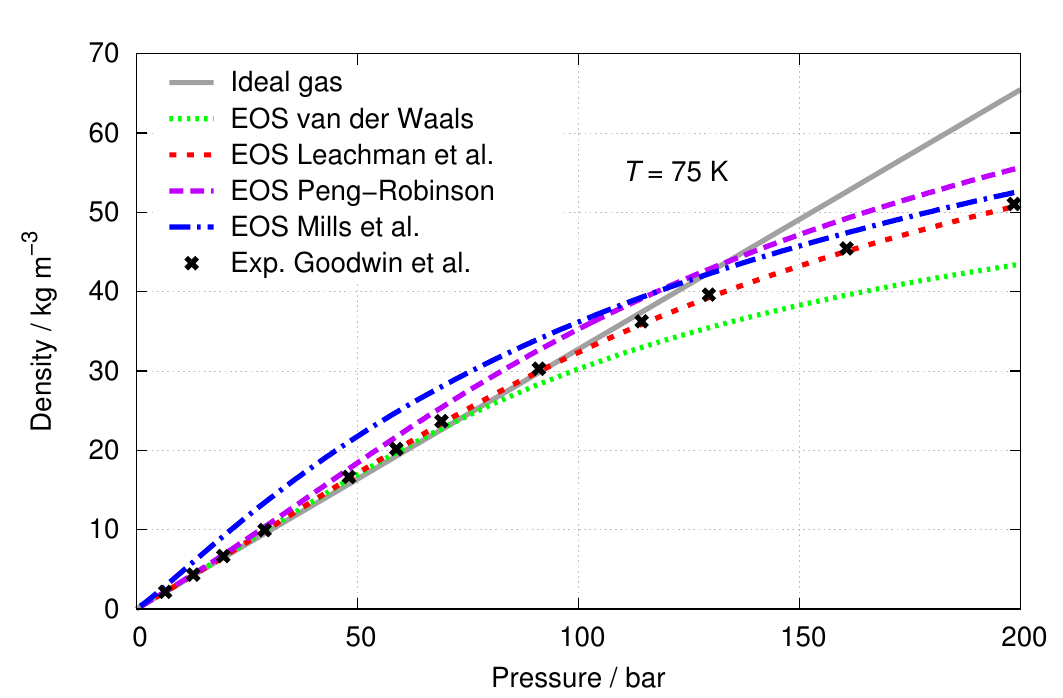}\caption{\label{fig:fig_02}Bulk H$_2$ densities at $T=75$
K calculated from different EOSs~\cite{Waals1873,Peng1976,Mills1977,Leachman2009}
are compared with experimental data of Goodwin et al.~\cite{Goodwin1963}.
The grey line represents the ideal gas.
}
\end{figure}

It is accepted text-book knowledge that the ideal gas approximation is
sufficient to treat the free hydrogen gas. Indeed, this statement is true
for a wide range of temperatures and pressures. In Figure~\ref{fig:fig_02},
various hydrogen reference isotherms at a temperature of 75 K are shown
and compared with the experimental data obtained by Goodwin et al.~\cite{Goodwin1963}.
Surprisingly, up to a pressure of 100 bar, the ideal gas isotherm is closer
to experiment than the van-der-Waals isotherm obtained from parameters
as published by the National Institute of Standards (NIST)~\cite{NIST},
which underestimates hydrogen densities starting from pressures of 70 bar.
At a pressure of 200 bar, those two isotherm approximations are unsuitable,
as the ideal gas equation overestimates the hydrogen density by 30\%, while
the real gas isotherm underestimates it by $\sim$20\%. For lower
temperatures, those differences are expected to get even worse.

Motivated by this deficiency of simple model isotherms, various groups
have parametrized hydrogen isotherms from experimental data. We have compared
data obtained from parametrized EOS by Mills et al.~\cite{Mills1977},
Peng and Robinson~\cite{Peng1976} and Leachman et al.~\cite{Leachman2009}.
The former two parametrizations overestimate the
hydrogen density at 75 K in the entire range of pressures. In particular,
the isotherm of Mills et al. performs rather poorly at pressures around
50 bar. At high pressure, however, they agree much better with the experimental
data than the ideal or van-der-Waals gas isotherms. Excellent agreement
with experiment is achieved by the EOS proposed by Leachman and coworkers~\cite{Leachman2009}.
At 75 K, the resulting isotherm matches the experiments
of Goodwin et al.~\cite{Goodwin1963} perfectly. Therefore, we have implemented
this EOS into our QLDFT code and use it for this and forthcoming studies
that aim at temperatures around 77 K. At this point, it is important to
note that at lower temperatures, ortho- and para-hydrogen behave differently,
thus the performance of the EOS should be reassessed for the particular application
that is examined. We further want to stress that even at low pressure applications,
the internal pressure at strong adsorption sites is typically much higher
than the external pressure. Thus, the EOS must be valid for a wide range
of pressures if hydrogen adsorption problems are to be studied.

State-of-the-art atomistic methods to treat gas adsorption (GCMC, QLDFT)
consider the situation that a significant part of the problem is the treatment
of the interactions between the adsorbed guest molecules. Indeed, the accurate
treatment of the interaction between hydrogen molecules is quite involved.
In fact, H$_2$ molecules are finite-size particles with a
directional quadrupole moment, and many-particle contributions do exist
\cite{Martinez-Mesa2012}. In this work, H$_2$ molecules
are modeled as single shapeless particles (that is, points) that are subject
of pairwise interactions. For the interaction between these particles,
we employ the classical (BO) potential of Diep and Johnson (DJ), that has
been obtained on grounds of high-level ab initio theory with extrapolation
to the basis set limit and correlation contributions~\cite{Diep2000}.
The distance-dependent term is parametrized in a Lennard-Jones (LJ) potential,
while the orientations of the hydrogen molecules with respect to each other
is taken into account by a series expansion in spherical harmonics. We
employ only the leading term of the DJ series, what is consistent with
the shapeless particle approximation of the H$_2$ molecule.
It is important to note that this potential reflects the BO potential and
hence does not include nuclear quantum effects, making it particularly
suitable for using it within QLDFT. 

\begin{figure}[h]
\centering{}\includegraphics[width=0.45\textwidth]{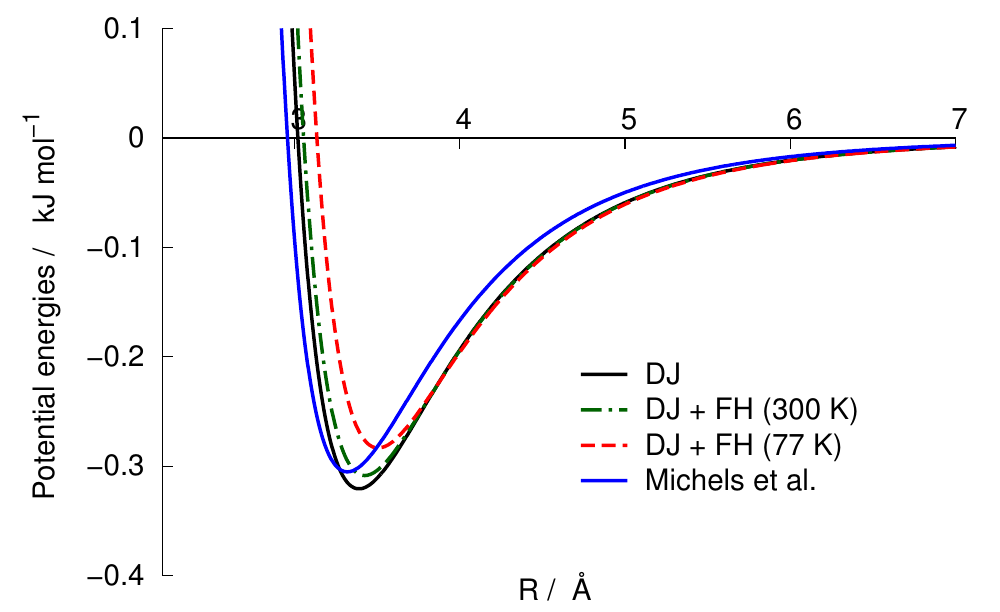}
\caption{\label{fig:fig_03}Comparison of classical and Feynman-Hibbs quantum corrected
potentials of H$_2$-H$_2$ interactions as function of the distance (R) between the centers of mass of the H$_2$ molecules.}
\end{figure}

A semi-classical alternative to the BO potential has been suggested by
Michels et al.~\cite{Michels1960}, who adjusted LJ parameters to match
the second virial coefficient, thus reflecting experimental $PVT$ data
of a temperature range of 98--423 K. We have compared both of the H$_2$-H$_2$
interaction potentials in Figure~\ref{fig:fig_03}. Michels et al.'s data
is close to the DJ potential at repulsion-dominated ranges, while the long-range
interactions are closer to the FH-corrected DJ curve.

\begin{figure}[h]
\centering\includegraphics[width=0.45\textwidth]{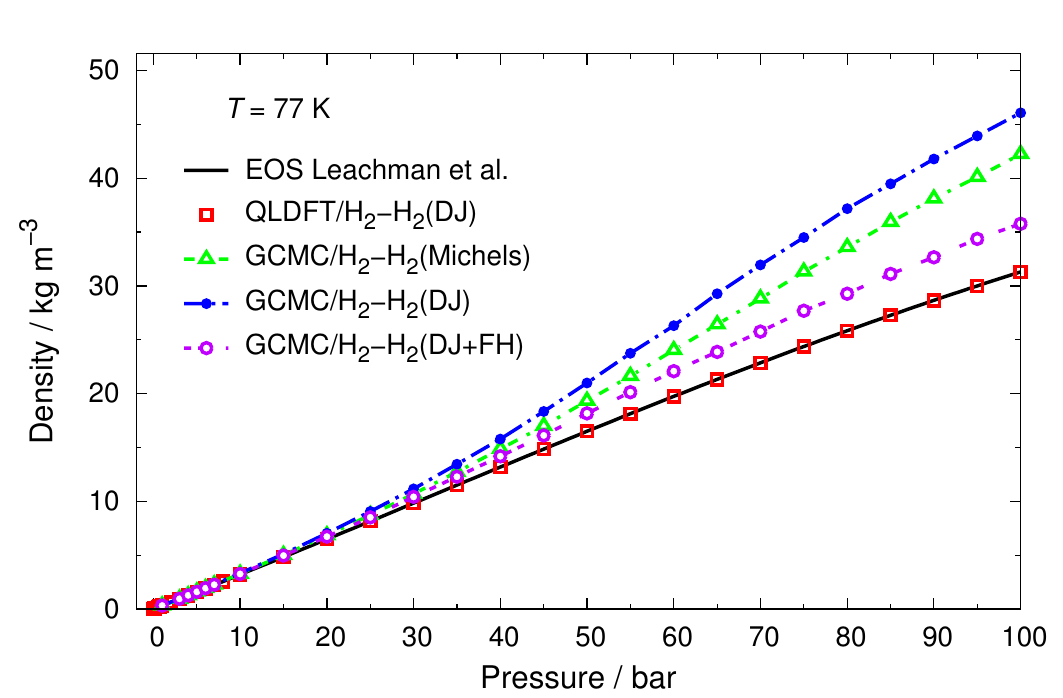}
\caption{\label{fig:fig_04} Real gas densities obtained from QLDFT and GCMC are
compared with the EOS of Leachman et al.~\cite{Leachman2009}}
\end{figure}

We have applied GCMC simulations to describe the free hydrogen gas, studied
at 77 K and in the pressure range of 0--100 bar. In Figure~\ref{fig:fig_04},
these calculations are benchmarked against our reference EOS by Leachman
et al.~\cite{Leachman2009}. For the classical treatment (GCMC with BO
DJ potential), the hydrogen density is strongly overestimated. Already
at pressures of 30 bar, it expects a 14\% higher hydrogen density, and
at 100 bar, the overestimation is 47\%. The situation improves for the
semi-classical quantum treatment, where the best performance is obtained
for the FH-corrected DJ potential. Still, a GCMC calculation employing
the FH-corrected DJ potential overestimates the H$_2$ density
by 14\% at 100 bar. On the other hand, by construction QLDFT results match
Leachman et al.'s isotherm perfectly. We conclude that treatment of the
free hydrogen gas using pairwise semi-classical potentials results in a
significant overestimation of the hydrogen density at higher pressures.

\section{Host-guest interaction potentials}

The host material acts via the external potential on the adsorbed molecules.
Thus, the external potential determines the adsorption properties of the
metal-organic framework. 
The repulsive part of the external potential defines
the inaccessible volume inside the framework and thus its inner surface.
The surface is rather sharply defined as the potential energy curve is
very steep at this region. The attractive part acts on the adsorbed gas
like a local pressure enhancement and thus increases the hydrogen density
at the surface. For a correct description of the adsorbed H$_2$
fluid the quality of the interaction potential is crucial. However, due
to fortuitous error compensations, calculated thermodynamic properties
may be close to experiment also for rather inaccurate external potentials.

The interaction of H$_2$ with the MOF framework is governed
by London dispersion interactions (LDI) that are known to be difficult
to be described at the quantum-mechanical level and at low computational cost. 
The state-of-the-art is to employ classical two-body force
field with an attractive r$^{-6}$ component for the long-range contribution,
and the repulsive part either described as repulsive r$^{-12}$ (Lennard-Jones)
or by an exponential (Buckingham) term. If the adsorption calculation includes
quantum effects as in QLDFT or PI-GCMC, then the external potential must
be strictly classical, that is, it needs to resemble the BO surface. If,
on the contrary, a classical adsorption calculation is pursued, as in GCMC,
LDFT or MD, then quantum effects could be treated semi-classically, either
by adding the Feynman-Hibbs correction to the potential, or by parametrizing
the FF such that experimental data are matched in the validation calculations.
Here, we will only consider FFs that are designed to predict the BO surface
of MOFs, thus eliminating empirical influences as much as possible. 

\begin{figure*}
\centering\includegraphics[width=0.9\textwidth]{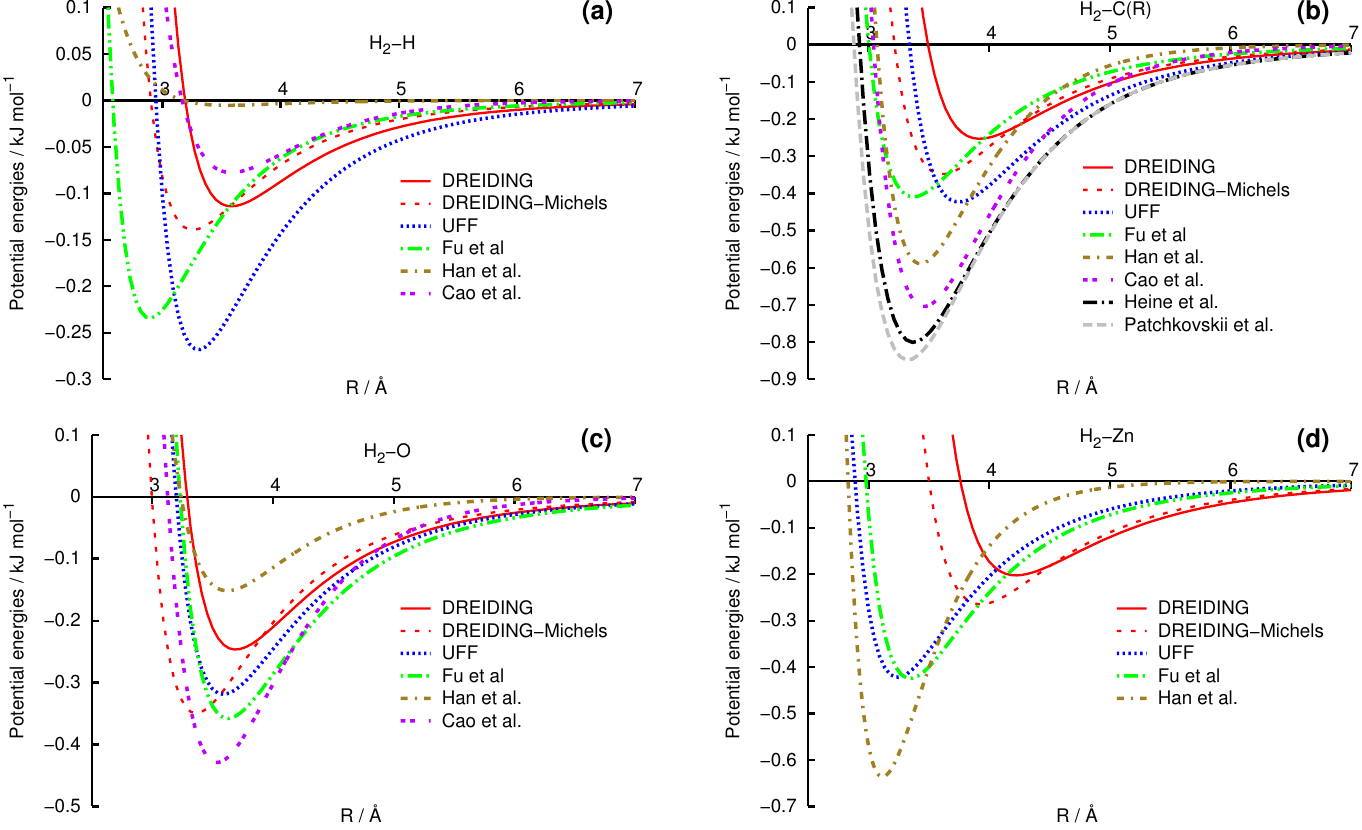}
\caption{\label{fig:fig_05}Typical interaction potentials between a hydrogen
molecule and framework atoms: H (a), C (b), O (c) and Zn (d) in the commonly
used DREIDING~\cite{Mayo1990} and UFF~\cite{Rappe1992a} force fields, compared with ab initio parameterized
force fields by Cao et al.~\cite{Cao2009}, Han et al.~\cite{Han2007a},
Fu et al.~\cite{Fu2009}, Heine et al.~\cite{Heine2004a} and Patchkovskiii et al.~\cite{Patchkovskii2005}.
R is the distance between the center of mass of the  H$_2$ and the framework atom}
\end{figure*}

\begin{figure*}
\centering\includegraphics[width=0.9\textwidth]{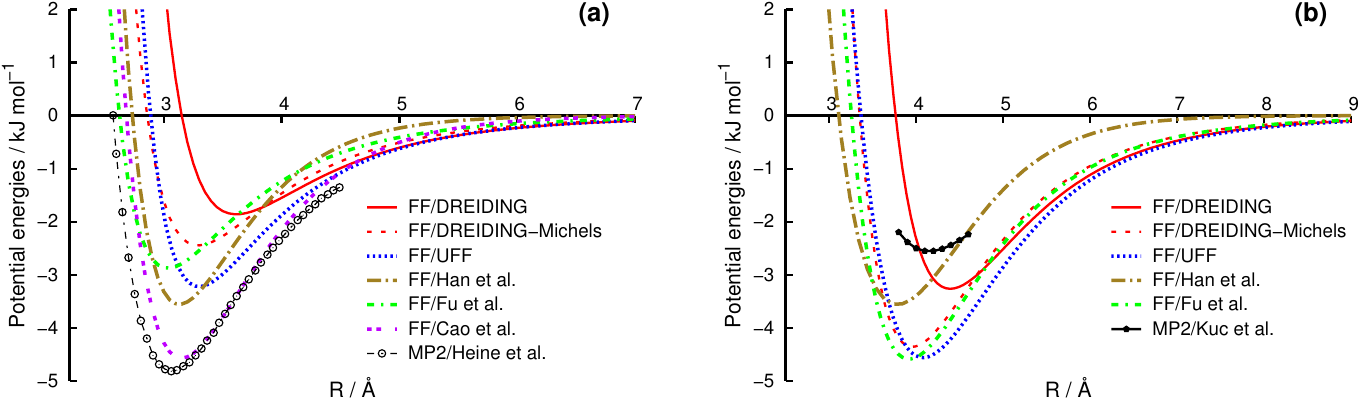}
\caption{\label{fig:fig_06}Potential energy curves of H$_2$ interacting with
Benzene and Zn$_4$O(HCO$_2$)$_6$
molecules, determined by different force fields, are compared with MP2 results:
(a) a spherical H$_2$ molecule moved along a line
perpendicular to the benzene plane and (b) a spherical H$_2$
molecule moved along the Zn--O axis in the $\alpha$ site of the Zn$_4$O(HCO$_2$)$_6$
cluster.  MP2 data points of Heine et al.~\cite{Heine2004a} and Kuc et al.~\cite{Kuc2008a}
are for diatomic H$_2$ molecules, where H$_2$ molecules are oriented parallel to the
axis of translation. R is the distance between the centers of mass of the molecules.}
\end{figure*}

In Figure~\ref{fig:fig_05}, we compare various interaction potentials
of H$_2$ with atom types that are relevant for MOFs. Already
at the first glance it is evident that these potentials are significantly
different. They disagree with the intercept with the energy axis -- that
is the position that roughly defines the internal surface, and with the
magnitude of the attraction of the respective atomic center with adsorbed
H$_2$. It is indeed surprising that many of them have been
applied to MOFs, often with comparable conclusions. As aromatic carbon
atoms provide a significant part of the LDI in MOFs, we will discuss here
the H$_2$-C potential in more detail. The only parameters
that have explicitly been developed for MOFs are those of Fu et al.~\cite{Fu2009}
and Han et al.~\cite{Han2007a}. Those are less attractive compared to
parameters that have been developed for all-carbon systems, e.g. those
of Patchkovskii et al.~\cite{Patchkovskii2005} and of Heine et al.~\cite{Heine2004a}.
Cao et al.'s~\cite{Cao2009} parameters have been optimized for covalent-organic
frameworks on the basis of MP2 calculations and also belong to the group
of strongly attractive FFs. On the other hand, the generic FFs (UFF~\cite{Rappe1992a},
DREIDING~\cite{Mayo1990}, and DREIDING with replaced H-H interactions
following Michel's pure gas isotherm~\cite{Michels1960}) are
less attractive, which is expected as they should include quantum effects.
In the FF of Han et al.~\cite{Han2007a} it is evident that no attraction
is seen in the H$_2$-H potential, which is compensated
by the stronger attractive H$_2$-C interaction. A similar effect
is present in the H$_2$-Zn and H$_2$-O interactions,
where Han et al.'s FF is strongly attractive to Zn, but less attractive
to O in comparison to Fu et al.'s~\cite{Fu2009} counterpart. For the present
work we have selected Fu et al.'s FF as it is constructed on the basis
of a large set of ab initio calculations. Benchmark calculations on benzene
and the Zn$_4$O(HCO$_2$)$_6$
cluster model of the $\alpha$ site in MOF-5 show that the overall performance
of both FFs are in much closer agreement than one would expect from the
inspection of the individual H$_2$-atomic components (see
Figure~\ref{fig:fig_06}).

It is hard to assess the performance of the different FFs: ab initio
calculations with converged results are difficult for systems that are
representable models approximating MOF building blocks. Moreover, the FFs
treat H$_2$ as shapeless particle and thus incorporate implicitly
the rotational degrees of freedom. Thus, we will validate the performance
of the FF by Fu et al.~\cite{Fu2009} on the basis of the system that is
most-widely studied, MOF-5, also known as IRMOF-1.

\section{Hydrogen adsorption in IRMOF-1: the role of quantum effects}

\begin{figure}[H]
\centering{}\includegraphics[width=0.45\textwidth]{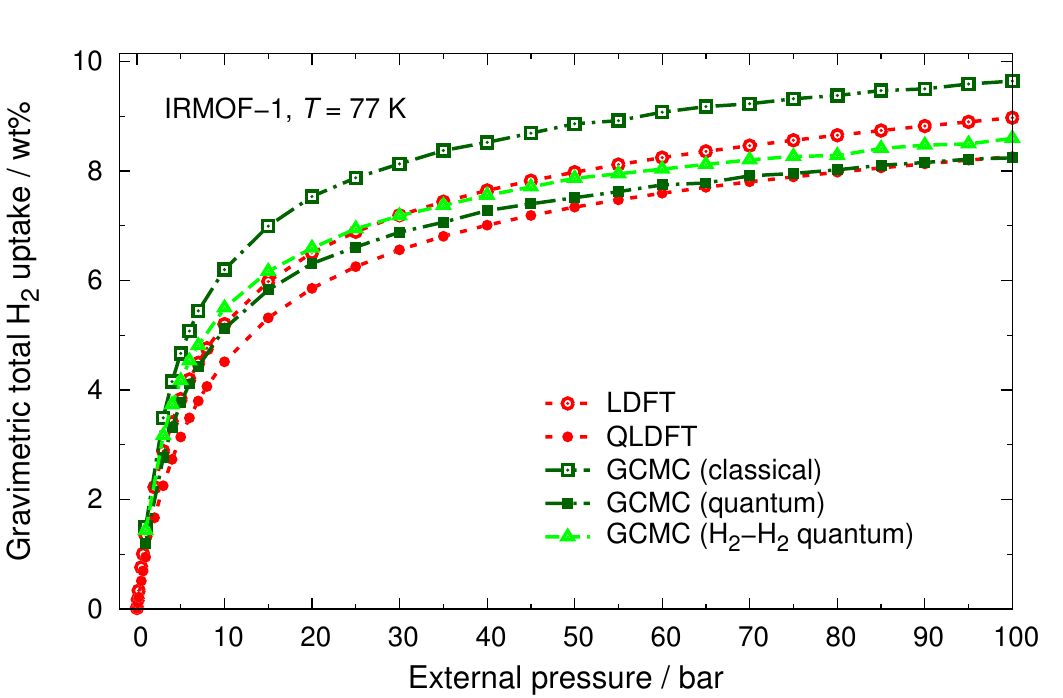}\caption{\label{fig:fig_07}Comparison of gravimetric total adsorption isotherms
of IRMOF-1 at $T=77$ K obtained from classical and quantum calculations
using the same interaction potentials and fugacity data.}
\end{figure}

Before validating our computational approach, we will first compare the
importance of quantum effects on the H$_2$ adsorption in
IRMOF-1. Quantum effects are totally absent in GCMC calculations if the
potential resembles the BO potential energy surface, that means, for any
potential that has been derived by electronic structure calculations, including
the one of Fu et al.~\cite{Fu2009}, but also those of Han et al.~\cite{Han2007a,Han2008a}
and the COF-FF of Cao et al.~\cite{Cao2009}. Therefore, not surprisingly,
the classical GCMC isotherm predicts the strongest adsorption capacity
(see Figure~\ref{fig:fig_07}). Inclusion of the semi-classical FH correction,
both for the H$_2$-H$_2$ as well as for the
H$_2$-host potentials, results in a substantial reduction
of hydrogen adsorption by $\sim$17\% at 77 K for pressures of
20 bar and above. A purely classical LDFT calculation is not possible within our implementation.
Even though if the kinetic energy operator
is absent in the Hamiltonian, a fair part of this effect is accounted for
in the XC potential that refers to the experimental free H$_2$
isotherm. The significantly lower adsorption of classical LDFT compared
to GCMC suggests that quantum effects in the adsorbed H$_2$
fluid are significant for higher pressures and should not be neglected.
This is confirmed by GCMC calculations where the H$_2$-H$_2$
interaction potential is subjected to the semi-classical FH correction
(see Figure~\ref{fig:fig_07}). Indeed, these GCMC results are close to LDFT.
Remaining differences are due to many-particle effects in the H$_2$
fluid and the approximate nature of the FH correction. 

QLDFT intrinsically accounts for quantum effects by explicit inclusion
of the kinetic energy in the Hamiltonian operator. Here, we should note
that this inclusion is not exact and follows a simple model, namely by
assuming a pseudo one-particle wave function that is obtained as square
root of the number density. Physically, this means that all H$_2$
molecules (bosons) are collapsed in the ground state. Thus, this approximation
works best for low temperatures, and it has indeed been shown that it performs
very well for liquid N$_2$ temperature (77 K)~\cite{Walther2013}.
Explicit treatment of quantum effects results again in lowering the H$_2$
adsorption density. However, the reduction is less compared to the difference
between GCMC calculations with and without quantum effects (see Figure~\ref{fig:fig_07}),
due to the fact that quantum effects inside the hydrogen fluid are implicitly considered within classical LDFT.
Finally, it should be noted that QLDFT and GCMC calculations with quantum
effects differ in the area of moderate pressures, mainly between 10 and
20 bar. As simulation model and interaction potential are identical, and
as we can exclude numerical issues for this difference, its origin can
either be the presence of many-body effects or the explicit quantum nature
of QLDFT, or contributions of both effects.

\begin{figure*}
\centering{}\includegraphics[width=0.9\textwidth]{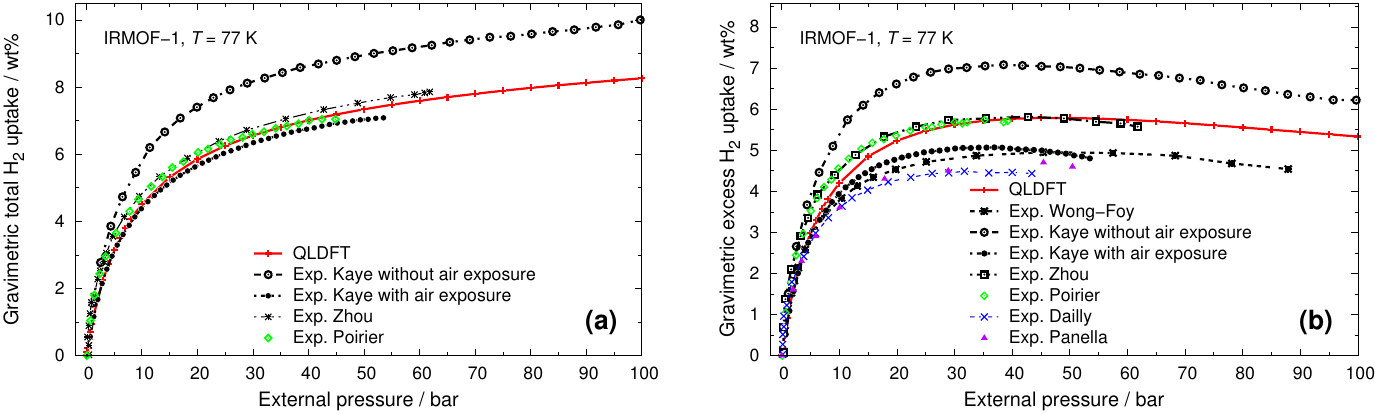}\caption{\label{fig:fig_08}Gravimetric total (a) and excess (b) adsorption isotherms
of IRMOF-1 at $T=77$ K obtained from QLDFT calculations are compared with
experiments~\cite{Zhou2007a,Kaye2007,Wong-Foy2006,Panella2006a,Poirier2008}. }
\end{figure*}

The QLDFT isotherm for IRMOF-1 is compared with available experimental
data from the literature. Our results show that the total gravimetric adsorption
isotherm predicted by QLDFT is in excellent agreement with those of Kaye
et al.~\cite{Kaye2007}(after air exposure), Zhou et al.~\cite{Zhou2007a},
and Poirier et al.~\cite{Poirier2008} (see Figure~\ref{fig:fig_08}). The
gravimetric H$_2$ uptake of IRMOF-1 before air exposure is
reported to be higher~\cite{Kaye2007}, possibly reflecting unsaturated
defects in the material. For the calculation of excess isotherms we have
subtracted the non-accessible part from the MOF volume. This has been
achieved by determining the space where the interaction potential is strongly
repulsive (we take the value of 5000 K as cutoff). The resulting QLDFT
gravimetric excess isotherm is in close agreement with the experimental
data of Zhou et al.~\cite{Zhou2007a} and Poirier et al.~\cite{Poirier2008},
and somewhat higher than the experiments of Kaye et al. (after air exposure)
\cite{Kaye2007} and of Wong-Foy et al.~\cite{Wong-Foy2006}. We conclude
that the QLDFT approach in conjunction with the FF parameters of Fu et
al.~\cite{Fu2009} satisfactorily describes the hydrogen adsorption in
this material. In the remainder, we apply this method to the series of
IRMOFs, and to MOF-177. These structures incorporate similar building units, thus, we expect them to be well-described by our methodology. 

\section{Hydrogen adsorption in the IRMOF series and MOF-177}

\begin{figure*}
\centering\includegraphics[width=0.9\textwidth]{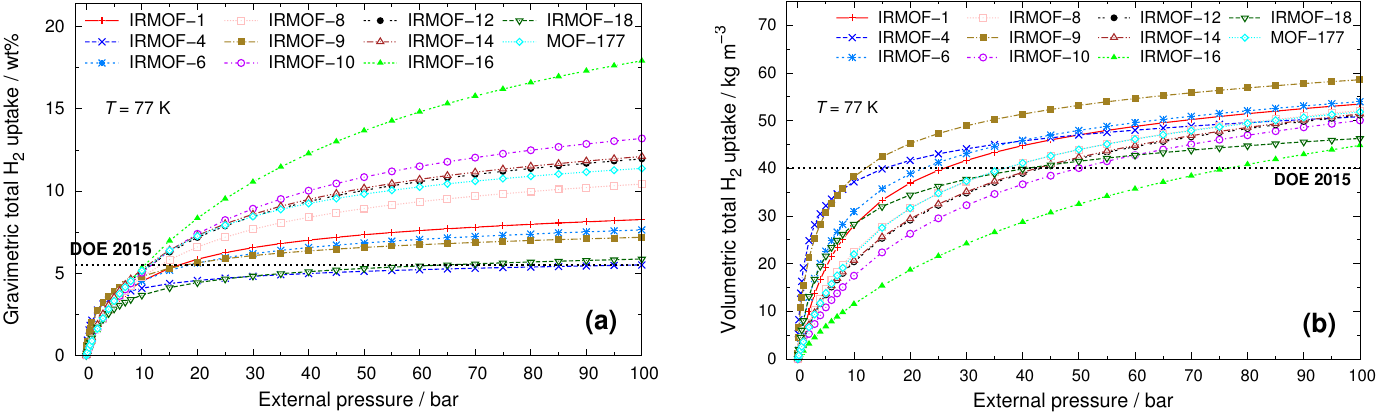}\caption{\label{fig:fig_09}Predicted gravimetric (a) and volumetric (b) total H$_2$
adsorption isotherms of MOFs at $T=77$ K. }
\end{figure*}

After validating our computational approach for IRMOF-1, we apply QLDFT
within the same setup as for IRMOF-1 to assess hydrogen adsorption properties
at $T=77$ K on a series of IRMOFs (IRMOF-n with n = 1, 4, 6, 8, 9, 10,
12, 14, 16, 18) and MOF-177. In Table~\ref{tab:table_1}, we list the total
gravimetric and volumetric adsorption capacities at 20, 40, 60, 80 and
100 bars, and the excess adsorption at saturation condition. Full isotherms
for $P=0-100$ bars are plotted in Figure~\ref{fig:fig_09}. 

Among all the MOFs studied here, at $77$ K, IRMOF-16 shows greatest gravimetric
hydrogen uptake (17.9 wt\% at $100$ bar). This structure possesses an
extraordinarily large surface area of 6063 m$^2$ g$^{-1}$,
and 84\% of its volume is accessible to the H$_2$ molecules.
However, the large free volume, coupled with a low crystal density
of 0.21 g cm$^{-3}$, results in the fact that IRMOF-16 shows
the lowest volumetric hydrogen uptake of all MOFs investigated here. A
similar trend has been observed for other MOFs, where materials with high
gravimetric uptake typically possess low volumetric capacities. Our results
show that among the materials considered here the doubly interpenetrated IRMOF-9 framework is the best hydrogen
adsorbent on the volumetric basis. 

\begin{table*}
\scriptsize{}
\setlength{\tabcolsep}{.4em}
\renewcommand{\arraystretch}{1.0}
\caption{Structural and adsorption properties of MOF systems studied in this work }
\label{tab:table_1} %
\begin{tabular}{lccccccccccccccc}
\toprule 
 & Free vol. & SASA  & \multicolumn{2}{c}{Total at 20 bar} & \multicolumn{2}{c}{Total at 40 bar} & \multicolumn{2}{c}{Total at 60 bar} & \multicolumn{2}{c}{Total at 80 bar } & \multicolumn{2}{c}{Total at 100 bar} & \multicolumn{3}{c}{Excess at saturation}\\
\cmidrule(r){4-5}\cmidrule(r){6-7} \cmidrule(r){8-9}\cmidrule(r){10-11}\cmidrule(r){12-13} \cmidrule(r){14-16}
  &  \%  & m$^2$g$^{-1}$  & wt\%  & kg m$^{-3}$  & wt\%  & kg m$^{-3}$  & wt\%  & kg m$^{-3}$  & wt\%  & kg m$^{-3}$  & wt\%  & kg m$^{-3}$  & P/bar  & wt\%  & kg m$^{-3}$\\
\midrule 
IRMOF-1  & 64.0  & 3799  & 5.9  & 36.9  & 7.0  & 44.7  & 7.6  & 48.8  & 8.0  & 51.4  & 8.3  & 53.5  & 45  & 5.8  & 36.5 \\
IRMOF-4  & 39.6  & 2170  & 4.6  & 41.7  & 5.0  & 45.7  & 5.2  & 48.0  & 5.4  & 49.6  & 5.5  & 50.8  & 45  & 4.4  & 40.5 \\
IRMOF-6  & 58.8  & 3467  & 5.6  & 38.9  & 6.6  & 45.9  & 7.1  & 49.6  & 7.4  & 52.1  & 7.6  & 54.0  & 45  & 5.5  & 38.3 \\
IRMOF-8  & 70.7  & 4436  & 6.6  & 31.6  & 8.4  & 41.1  & 9.4  & 46.2  & 10.0  & 49.6  & 10.4  & 52.1  & 55  & 6.7  & 32.3 \\
IRMOF-9  & 55.8  & 3746  & 5.6  & 45.2  & 6.4  & 51.4  & 6.7  & 54.6  & 7.0  & 56.9  & 7.2  & 58.6  & 45  & 5.5  & 44.1 \\
IRMOF-10  & 77.4  & 4923  & 7.4  & 26.3  & 10.0  & 36.6  & 11.5  & 42.8  & 12.5  & 46.9  & 13.2  & 50.0  & 60  & 7.7  & 27.5 \\
IRMOF-12  & 74.3  & 4939  & 7.2  & 29.1  & 9.4  & 38.9  & 10.6  & 44.5  & 11.4  & 48.3  & 12.0  & 51.1  & 55  & 7.4  & 29.9 \\
IRMOF-14  & 75.0  & 4783  & 7.3  & 29.4  & 9.5  & 39.2  & 10.7  & 44.8  & 11.5  & 48.6  & 12.1  & 51.4  & 55  & 7.4  & 30.0 \\
IRMOF-16  & 84.1  & 6063  & 8.3  & 18.7  & 12.3  & 28.7  & 14.8  & 35.7  & 16.6  & 40.8  & 17.9  & 44.8  & 70  & 8.6  & 19.2 \\
IRMOF-18  & 48.2  & 2571  & 4.4  & 34.4  & 5.1  & 39.9  & 5.5  & 42.8  & 5.7  & 44.7  & 5.9  & 46.2  & 45  & 4.3  & 33.6 \\
MOF-177  & 70.3  & 4901  & 7.3  & 31.6  & 9.2  & 41.1  & 10.3  & 46.1  & 10.9  & 49.4  & 11.4  & 51.8  & 55  & 7.4  & 32.3 \\
\bottomrule
\end{tabular}
\end{table*}

In this study, we did not predict room temperature adsorption capacities.
Among others, Frost et al.~\cite{Frost2007} already pointed out that the
strength of the adsorption potential of these materials is not high enough
to adhere sufficient amount of H$_2$ molecules that fulfil
the requirements suggested by the US Department of Energy (DOE) at room temperature~\cite{DOE2009}. Fu et al.~\cite{Fu2009}
calculated the room temperature adsorption for the majority of the MOFs studied
in this work and confirmed very poor room temperature adsorption. However,
our results show that most of the MOFs can even surpass the DOE targets
\cite{DOE2009} at technologically plausible liquid N$_2$
temperature (see Figure~\ref{fig:fig_09}) at moderate pressures. This
may indeed be a viable alternative to high-pressure tanks, as a more flexible
tank design would be possible. 

\begin{figure*}
\centering\includegraphics[width=0.9\textwidth]{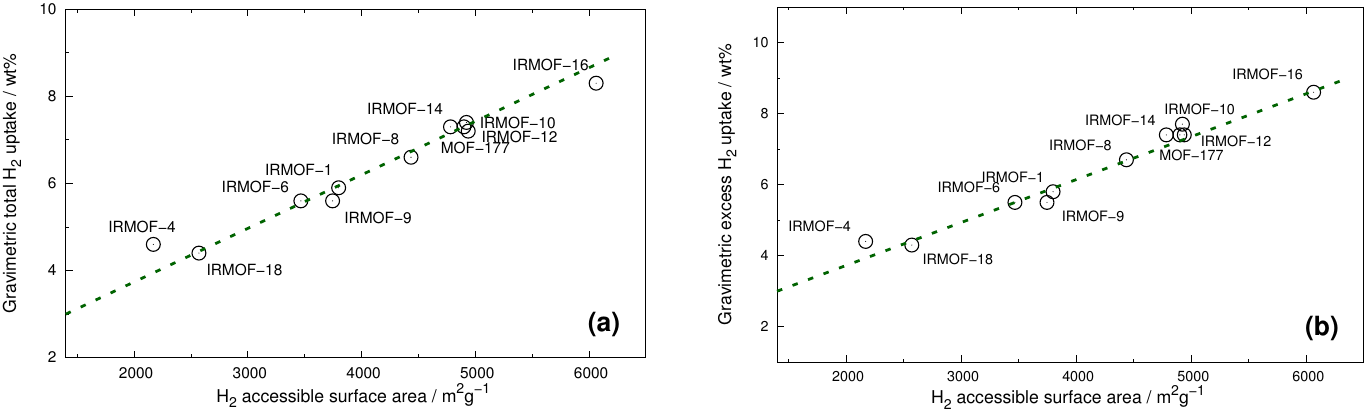}\caption{\label{fig:fig_10}Relationship between H$_2$ accessible
surface area and gravimetric total H$_2$ adsorption at $P=20$
bar (a), and gravimetric excess H$_2$ adsorption at saturation
pressures (b).}
\end{figure*}

\begin{figure}[h]
\centering\includegraphics[width=0.45\textwidth]{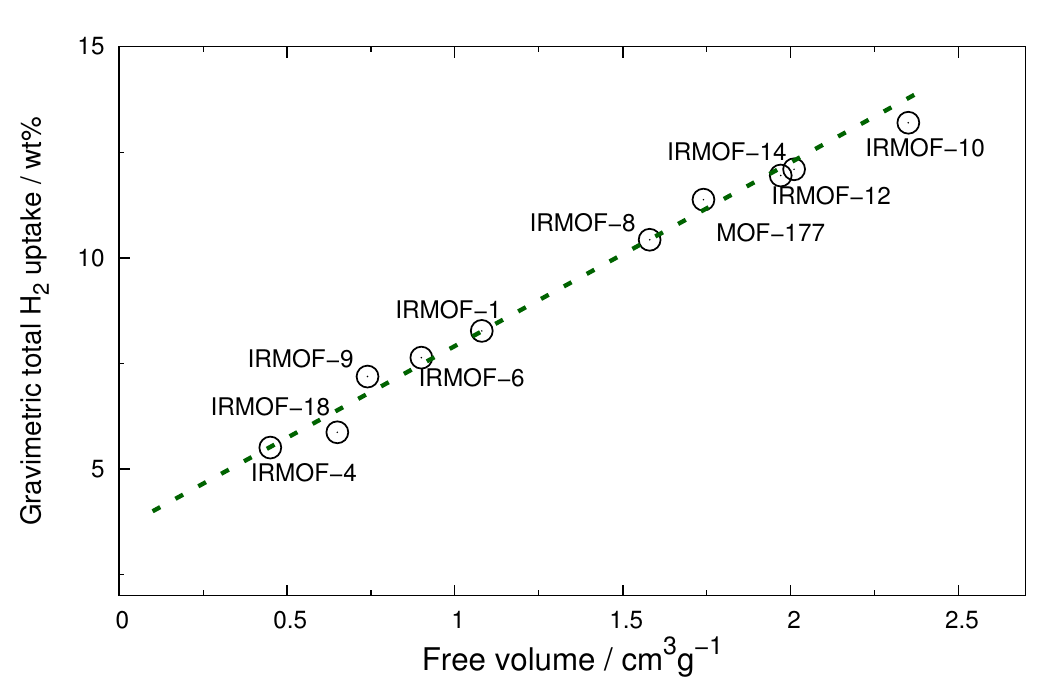}\caption{\label{fig:fig_11}Relationship between H$_2$ accessible
free volume and total amount of adsorbed H$_2$ at
$P=100$ bar.}
\end{figure}

\begin{figure*}
\centering\includegraphics[width=0.9\textwidth]{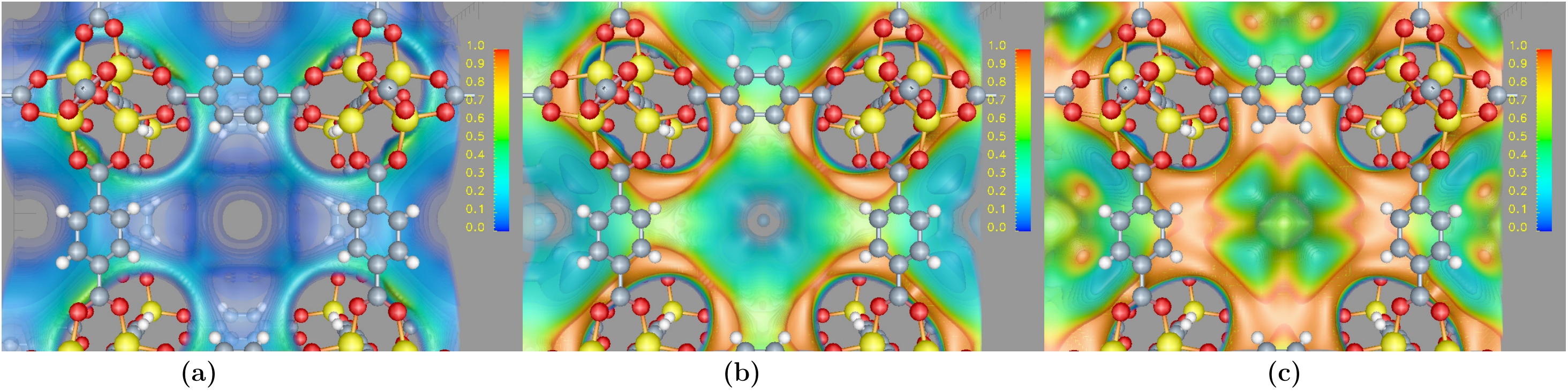}
\caption{\label{fig:fig_12}Normalized density profiles of adsorbed H$_{2}$ inside
the unit cell of IRMOF-1 at $T=77$ K, and $P=$1 bar (a), 10 bar (b) and
100 bar (c).}
\end{figure*}

In Figure~\ref{fig:fig_10}, we have plotted the gravimetric total uptakes
at 20 bar and excess uptakes at saturation condition against the solvent
accessible surface area of corresponding MOFs. Our results show that adsorption
capacities at mid pressure regime directly correlate with the solvent
accessible surface area and the correlation exists up to the saturation
regime (see Figure~\ref{fig:fig_10}). We have also found a linear correlation
between pore volume and gravimetric uptake at 100 bar (see Figure~\ref{fig:fig_11}).
Similar correlations have been reported by Frost et al.~\cite{Frost2006}.
In order to better illustrate these correlations, we plotted the evolution
of adsorbed H$_2$ density profile inside the IRMOF-1 pore at
different pressure loadings (see Figure~\ref{fig:fig_12}). Due to the
attractive host-guest interaction potentials, the first monolayer of adsorbed
hydrogen would stay about one molecular radius apart from the sorbent surface.
The formation of few more monolayers is possible, depending on the behavior
of how  the potential energy diminishes at larger distances. In general,
the hydrogen molecules stay close to the surface and rest of  the pore
remains empty at low pressure. For higher pressures, hydrogen molecules
gradually start occupying the pore interior (see Figure~\ref{fig:fig_12}).

\subsection*{Comparison with experiment and other theoretical studies}

To our knowledge, this is the first time that the hydrogen adsorption in
a class of MOFs has been investigated using as little empirical tuning
as possible. Indeed, to date the hydrogen adsorption properties of the
IRMOF series and of MOF-177 have been widely studied by experiments~\cite{Rowsell2004,Rowsell2006,Wong-Foy2006,Dailly2006,Furukawa2007,Panella2006a,Panella2008,Poirier2008,Zhou2007a,Hirscher2010}
and by computer simulations~\cite{Yang2005,Jung2006,Han2008a,Han2007,Han2007a,Frost2006,Ryan2008,Fu2009,Garberoglio2005,Klontzas2008a}.
In both cases, it is difficult to arrive at a conclusion on the performance
of a particular MOF. On the experimental side, synthesis and transport
conditions, defect concentration and experimental details measuring the
hydrogen adsorption lead to a rather wide range of observed adsorption
capacities. On the other hand, in the majority of the theoretical works,
the employed potential models and the computational approaches were calibrated
against low pressure (0--1 bar) experimental data. This has led to poor
agreement in a wider pressure range~\cite{Keskin2009}. 

One should also keep in mind that the conversion of simulation results
to  the experimental equivalent (i.e., total to excess) or vice versa,
required knowledge about the  H$_2$ accessible volume of
the frameworks pore and free hydrogen densities at certain thermodynamic
condition. This data is another source of uncertainty that can strongly
affect the results. Accepting these limitations, a quantitative agreement
of adsorption results might be challenging, but a qualitative agreement
(curvature of the isotherms) can be achieved~\cite{Frost2006,Ryan2008,Fu2009}.
Figure~\ref{fig:fig_08} exemplifies these facts: our QLDFT determined
isotherms of IRMOF-1 match with most of the experiments qualitatively,
but quantitative agreement is achieved for few of them only. 

In Table~\ref{tab:table_2}, we have compared the adsorption capacities
of the MOF systems determined from our QLDFT calculations with available
experimental and simulation data. Except for IRMOF-8, our predicted results
are in good agreement with literature data. In the worst case, IRMOF-8,
our excess isotherms significantly exceed the experimental results of Panella
et al.~\cite{Panella2008} and Dailly et al.~\cite{Dailly2006}, who have
reported that IRMOF-8 gets saturated at the 10--15 bar pressure range adsorbing
3.6 wt\% H$_2$, while our calculations indicate that IRMOF-8
can adsorb a maximum 6.7 wt\% at 55 bar. Independently, Garberoglio et
al.~\cite{Garberoglio2005} also predicted a saturation pressure similar
to ours, even with higher adsorption ($\sim$8.3 wt\%). GCMC
simulations of Fu et al.~\cite{Fu2009}, Frost et al.~\cite{Frost2006},
and Yang and Zhong~\cite{Yang2005} also predicted higher adsorption compared
to experiments. In addition, we have
thoroughly compared our QLDFT results with Fu et al.'s~\cite{Fu2009} Feynman-Hibbs
quantum corrected and Frost et al.'s~\cite{Frost2006} classical GCMC results
for the same MOF systems studied in this work. In terms of gravimetric
total uptakes at 80 bar, Fu et al's and Frost et al.'s consistently predict higher hydrogen uptake by 5--10\% and 12--19\%, respectively.

\begin{table*}
\scriptsize{}
\renewcommand{\arraystretch}{0.8}\caption{QLDFT determined H$_2$ adsorption results are compared
with experiments and GCMC data\textsuperscript{a}}
\label{tab:table_2}
\centering{}%
\begin{tabular*}{0.60\textwidth}{@{\extracolsep{\fill}}llllll}
\toprule 
 & \multicolumn{2}{c}{Excess uptake at 77 K} & \multicolumn{2}{c}{Total uptake at 77 K} & Method and Ref.\\
\cmidrule(r){2-3}\cmidrule(r){4-5} 
 & $P$ / bar & wt\% & $P$ / bar & wt\% & \\
\midrule
IRMOF-1 & 30{*} & 4.3 &  &  & Exp. [\onlinecite{Dailly2006}]\\
        & 45{*} & 4.7 &  &  & Exp. [\onlinecite{Panella2006a}]\\
        & 45{*} & 4.9 &  &  & Exp. [\onlinecite{Wong-Foy2006}]\\
        & 40{*} & 5.1\textsuperscript{b},7.1\textsuperscript{c} & 45 & 6.9\textsuperscript{b},8.8\textsuperscript{c} & Exp. [\onlinecite{Kaye2007}]\\
        & 45{*} & 5.8 & 45 & 7.4 & Exp. [\onlinecite{Zhou2007a}]\\
        & 45{*} & 5.8 & 45 & 7.2 & QLDFT\\
\midrule
IRMOF-4 &  &  & 80 & 5.1 & GCMC [\onlinecite{Frost2006}]\\
        &  &  & 80 & 5.4 & QLDFT\\
\midrule
IRMOF-6 & 45{*} & 4.6 &    &     & Exp. [\onlinecite{Wong-Foy2006}]\\
        & 45{*} & 5.5 & 80 & 7.4 & QLDFT\\
        &       &     & 80 & 8.3 & GCMC [\onlinecite{Frost2006}]\\
\midrule
IRMOF-8 & 10--15{*} & 3.6 &  &  & Exp. [\onlinecite{Dailly2006},\onlinecite{Panella2008}]\\
        & 10--15 & 4.3--5.1 &  &  & QLDFT\\
        & 55{*} & 6.7 & 80 & 10.0 & QLDFT\\
        & 50{*} & 8.3 & 80 & 11.9 & GCMC [\onlinecite{Garberoglio2005}]\\
        & 80 & 7.7 & 80 & 10.8 & GCMC [\onlinecite{Fu2009}]\\
        &  &  & 80 & 11.8 & GCMC [\onlinecite{Frost2006}]\\
\midrule
IRMOF-9 & 1 & 1.2 &  &  & Exp. [\onlinecite{Rowsell2006}]\\
        & 1 & 1.9 & 80 & 7.0 & QLDFT\\
        &  &  & 80 & 7.7 & GCMC [\onlinecite{Fu2009}]\\
\midrule
IRMOF-10 &  &  & 80 & 13.4 & GCMC [\onlinecite{Fu2009}]\\
         &  &  & 80 & 14.7 & GCMC [\onlinecite{Frost2006}]\\
         &  &  & 80 & 12.5 & QLDFT\\
\midrule
IRMOF-12 &  &  & 80 & 13.4 & GCMC [\onlinecite{Frost2006}]\\
         &  &  & 80 & 11.4 & QLDFT\\
\midrule
IRMOF-14 & 50{*} & 10 & 100 & 14.6 & GCMC [\onlinecite{Garberoglio2005}]\\
         & 55{*} & 7.4 & 100 & 12.1 & QLDFT\\
\midrule
IRMOF-16 &  &  & 80 & 18.0 & GCMC [\onlinecite{Fu2009}]\\
         &  &  & 80 & 19.7 & GCMC [\onlinecite{Frost2006}]\\
         &  &  & 80 & 16.6 & QLDFT\\
\midrule
IRMOF-18 & 1 & 0.9 &  &  & Exp. [\onlinecite{Rowsell2004}]\\
 & 1 & 1.0 & 80 & 5.7 & QLDFT\\
 &  &  & 80 & 5.5 & GCMC [\onlinecite{Frost2006}]\\
\midrule
MOF-177 & 52--66{*} & 6.8-7.1 & 72--75 & 10.0--10.3 & Exp. [\onlinecite{Furukawa2007}]\\
        & 69{*} & 7.0 & 78 & 10.2 & Exp. [\onlinecite{Wong-Foy2006}]\\
        & 55{*} & 7.4 & 75 & 10.0 & QLDFT\\
\bottomrule
\multicolumn{6}{l}{\textsuperscript{a}: some data points are extracted from the published
figures, consider a small deviation from the original. }\\
\multicolumn{6}{l}{\textsuperscript{b }and \textsuperscript{c}: sample prepared with and
without air exposure}\\
\multicolumn{6}{l}{{*} indicates the pressure at which the system reached its maximum excess
adsorption}\\
\end{tabular*}
\end{table*}

\section{Conclusion}

We have calculated the role of quantum effects on the adsorption of molecular
hydrogen in metal-organic frameworks. The inclusion of quantum effects
significantly reduces the hydrogen adsorption capacity at liquid N$_2$
temperature ($T=77$ K) compared to classical simulations, and therefore
it is crucial to include them in the calculation in order to make correct
predictions. They can be included either explicitly, e.g. within Quantized
Liquid Density-Functional Theory, or semi-classically in any classical
method of statistical mechanics as for example the Grand Canonical Monte-Carlo
approach. Both approaches are computationally competitive, QLDFT having
the advantage that it implicitly accounts for many-body effects via the
XC potential, while GCMC is suitable to treat hydrogen as diatomic species
beyond the shapeless particle approximation. It remains challenging to
have a satisfactorily working host-guest potential, first as two-body force
fields are intrinsically limited in their accuracy, and as high-quality
reference calculations are computationally demanding.

Our calculations are, however, in good agreement with experimental data.
At technologically applicable liquid N$_2$ temperature ($T=77$
K) and moderate pressure (20 bar or even less), many MOFs fulfil the requirements
of the US Department of Energy of 5.5 wt\% gravimetric and 40 kg m$^{-3}$
volumetric storage capacity, which however are defined for room temperature
storage. It should be noted, though, that the choice for the best MOF material
depends on the technological context: Those frameworks with exceptionally
high gravimetric storage capacity perform poorly in volumetric storage
and vice versa. Results also depend on the target pressure that is suitable
in the system. Thus, the implementation of a H$_2$ storage
facilities using MOFs are technologically possible, though economically
demanding.

Finally, we conclude that our QLDFT approach is an available and computationally
feasible alternative to popular classical or semi-classical approaches
of statistical mechanics that allows the explicit treatment of quantum
effects and implicitly includes many-body interactions in the adsorbed
hydrogen fluid. 
\begin{acknowledgments}
This work has been supported by the European Research Council (ERC-StG, GA 256962)
and Deutsche Forschungsgemeinschaft (GA HE 3543/7-2).
We thank Dr. A. Martínez-Mesa for providing the subroutine of Leachman et al.'s EOS. 
\end{acknowledgments}

\bibliography{H2-MOFs-JCP}

\end{document}